\documentclass[checkin,showpacs,aps,prb]{revtex4-1}
\usepackage{pifont}
\usepackage{amssymb}
\usepackage{datetime}
\usepackage{setspace}

\usepackage{amsmath,bm}
\usepackage[table]{xcolor}
\usepackage{graphicx}

\newcommand{\gc}[1]{{^{(#1)}}}
\newcommand{\gd}[2]{{^{(#1)}_{#2}}}

\newcommand{\bn}{\begin{enumerate}}
\newcommand{\en}{\end{enumerate}}
\newcommand{\ba}{\begin{eqnarray}}
\newcommand{\ea}{\end{eqnarray}}

\newcommand{\ub}{$\mu_B$}

\newcommand{\be}{\begin{equation}}
\newcommand{\ee}{\end{equation}}
\newcommand{\la}{\langle}

\newcommand{\ra}{\rangle}
\newcommand{\et}{{\it et al. }}

\newcommand{\bA}{{\bf A}}
\newcommand{\br}{{\bf r}}
\newcommand{\bp}{{\bf p}}
\newcommand{\bP}{{\bf P}}
\newcommand{\bD}{{\bf D}}
\newcommand{\bE}{{\bf E}}

\newcommand{\bS}{{\bf S}}
\newcommand{\bM}{{\bf M}}

\newcommand{\hp}{\hat{\bp}}

\newcommand{\nk}{n{\bf k}}

\newcommand{\mje}{mJ/cm$^2$}
\newcommand{\mj}{mJ/cm$^2$\ }

\newcommand{\bk}{{\bf k}}

\def\prl{{ Phys. Rev. Lett. }}

\begin{document}










\title{Nonlinear optical quantum theory of demagnetization in L1$_0$
  FePt and FePd}


\author{G. P. Zhang$^*$} \affiliation{Department of Physics, Indiana
  State University, Terre Haute, Indiana 47809, USA}

\author{Y. H. Bai} \affiliation{Office of Information
  Technology, Indiana State University, Terre Haute, Indiana 47809,
  USA}

 \author{Thomas F. George} \affiliation{Departments of Chemistry \&
   Biochemistry and Physics \& Astronomy \\University of
   Missouri-St. Louis, St.  Louis, MO 63121 }

\date{\today}

%

\begin{abstract}
  {It is now well established that a laser pulse can demagnetize a
    ferromagnet. However, for a long time, it has not had an analytic
    theory because it falls into neither nonlinear optics (NLO) nor
    magnetism. Here we attempt to fill this gap by developing a
    nonlinear optical theory centered on the spin moment, instead of
    the more popular susceptibility. We first employ group theory to
    pin down the lowest order of the nonzero spin moment in a
    centrosymmetric system to be the second order, where the
    second-order density matrix contains four terms of sum frequency
    generation (SFG) and four terms of difference frequency generation
    (DFG). By tracing over the product of the density matrix and the
    spin matrix, we are now able to compute the light-induced spin
    moment.  We apply our theory to FePt and FePd, two most popular
    magnetic recording materials with identical crystal and electronic
    structures. We find that the theory can clearly distinguish the
    difference between those two similar systems. Specifically, we
    show that FePt has a stronger light-induced spin moment than FePd,
    in agreement with our real-time ultrafast demagnetization
    simulation and the experimental results. Among all the possible
    NLO processes, DFGs produce the largest spin moment change, a
    manifestation of optical rectification.  Our research lays a solid
    theoretical foundation for femtomagnetism, so the light-induced
    spin moment reduction can now be computed and compared among
    different systems, without time-consuming real-time calculations,
    representing a significant step forward. }
\end{abstract}



 \maketitle

 \section{Introduction}

 \newcommand{\bb}{{\bf B}}
  \newcommand{\bs}{{\bf S}}


 \newcommand{\bq}{{\bf q}}
 \newcommand{\bQ}{{\bf Q}}

{ 
Using the light to change magnetic properties in semiconductors,
antiferromagnets and ferrites \cite{kurita1981} can be traced back to
several decades ago\cite{holzrichter1971}.  Using an
ultrafast laser pulse to demagnetize magnetic materials \cite{eric}
launched a new frontier, femtosecond magnetism, or femtomagnetism. The
pioneering work by Beaurepaire and coworkers inspired several decades
of intense investigations (see reviews
\cite{ourreview,rasingreview}).  The laser pulses can also switch
spins from one direction to another permanently
\cite{stanciu2007,jpcm24}. These discoveries are potentially
applicable to spintronics, with improved speed and efficiency (see
reviews \cite{rasingreview,ourbook,jpcm24}).}

However, in contrast to femtochemistry in molecules
\cite{zewail1988,zewail} and femtobiology in photoisomerization in
rhodopsins and yellow proteins
\cite{gai1998,schnedermann2016,mukamel,fayer,sundstrom2008},
femtomagnetism lacks a firm grounding at a quantitative level in
neither nonlinear optics (NLO) nor magnetism.  The traditional spin
wave theory \cite{kubler} is based on the Heisenberg spin exchange
model or Stoner's spin wave model without a light field
\cite{kittel,ashcroft,callaway,li,kubler}, where spin-wave excitation
(magnon) is at the center of demagnetization \cite{kittel,jap19}.  On
the other hand, NLO centers on the light generation, not the spin
moment change, and is formulated around different orders of
susceptibilities \cite{bloembergen,shen,butcher,boyd}.  {For
  instance, the traditional magneto-optics
  \cite{jmmm98,bennemann,oppeneer2004,np09} investigates
  how a magnetic field affects the generated light signal}.  The
inverse Faraday effect (IFE) \cite{shen} is probably the only
exception, but there is no guarantee for demagnetization instead for
magnetization. A prior study \cite{berritta2016} focused on the
helicit-dependent Faraday constants in simple $3d$ ferromagnets,
without real-time-dependent simulation, so this does not explain the
above ultrafast demagnetization experiments, where the spin moment is
always decreased. To the best of our knowledge, a nonlinear optical
theory that can yield a negative spin moment change does not exist.

In this paper, we aim to develop a nonlinear optical quantum theory
for demagnetization perturbatively, by focusing on the spin moment
change in ferromagnets.  We employ two centrosymmetric magnetic
materials, FePt and FePd, with similar crystal and electronic
structures.  We start with the symmetry analysis and find that the
first-order light-induced spin moment for a centrosymmetric system is
zero, and the lowest nonzero spin moment is the second order. This
sets the framework for our NLO theory. There are in total eight NLO
processes: four are sum frequency generation (SFG) and the other four
are difference frequency generation (DFG). SFGs contribute a tiny spin
change, but the contribution from DFGs is very large. Physically, DFGs
correspond to the optical rectification.  It is the competition
between two DFGs that leads to the negative spin moment. Under the
same fluence, FePt has a stronger spin moment reduction than FePd,
since the former has a stronger spin-orbit coupling.  To verify this
result, we carry out the time-dependent simulation of spin moment
change and find that dynamically, FePt demagnetizes more, regardless
of the laser photon energy, pulse duration, and fluence, fully
consistent with the experimental results by Iihama \et
\cite{iihama2016}.  Our study represents a serious attempt to
understand ultrafast demagnetization, without actual time-dependent
simulation.  Our theory will have a significant impact on future
research, by providing a means to compute and contrast different
materials at the quantitative first-principles level, greatly
enhancing accessibility and reproducibility to the broader research
community of ultrafast spintronics and all-optical spin switching.

The rest of the paper is arranged as follows. In Sec. II, we present
our theoretical formalism and the ground-state properties of FePt and
FePd. Section III is devoted to the nonlinear optical quantum theory
for spin response, where we start from the symmetry analysis, and then
move on to an analytic theory for the spin change. In Sec. IV, we
carry out the real-time simulation of ultrafast demagnetization under
laser excitation to realistically test the results of nonlinear
optical quantum theory. We conclude this paper in Sec. V. We provide a
detailed derivation of our main formulas in Appendix \ref{app}.
{Since our theory uses a good number of symbols
  which the reader may find  difficult to follow, we list them in
  our Table \ref{table0}}.

\section{Theoretical formalism}

There is no better example than FePt and FePd. They are among the most
studied materials for magnetic recording
\cite{cui2010,qin2017,lyu2024}, with a large magnetic anisotropy
\cite{burkert2005}.  The dependence of ultrafast demagnetization on
the Mn doping \cite{liu2020}, temperature and fluence \cite{xie2023}
have been thoroughly examined.  Yamamoto \et \cite{yamamoto2019}
investigated the ultrafast demagnetization at the Pt-edge.  Liu \et
\cite{liu2009} employed a single pump pulse to excite FePt, and found
that the pump reduces their Kerr hysteresis loop but not the
coercivity.  Shi \et \cite{shi2018} utilized two pump pulses of the
same fluence, and showed that the coercivity is reduced once the laser
fluence is above 4 \mje. This important result reveals a possible
onset for magnetic domain changes that are crucial for all-optical
spin switching.  All-optical spin switching was reported in FePt
nanoparticles \cite{john2017} and Cr- and Mn-doped FePt films
\cite{stiehl2024}.

Theoretically, we employ the state-of-the-art density functional
theory as implemented in the Wien2k code \cite{wien2k}. We
numerically solve the Kohn-Sham equation \be \left
[-\frac{\hbar^2\nabla^2}{2m_e}+V_{Ne}+V_{H}+V_{xc} \right
]\psi_{\nk}(\br)=E_{\nk} \psi_{\nk} (\br), \label{ks} \ee where the
terms on the left are the kinetic energy operator, the attraction
between the nuclei and electrons, the Hartree term, and the
exchange-correlation at the PBE level, respectively.  
  The spin-orbit coupling is included using a second-variational
  method in the same self-consistent iteration.  FePt and FePd share
similar structures and crystallize in a face-centered tetragonal
L1$_0$ structure \cite{laughlin2005}, with space group No. 123,
P4/mmm.  Figure \ref{fig1}(a) illustrates that Fe occupies two
inequivalent sites $1a(0,0,0)$ and $1c(\frac{1}{2},\frac{1}{2},0)$,
and Pt/Pd takes two equivalent sites
$(2e)(0,\frac{1}{2},\frac{1}{2}),(\frac{1}{2},0,\frac{1}{2})$, so
there are four atoms in the conventional tetragonal face-centered unit
cell. They also have similar lattice constants $a=b=3.859\ \rm \AA $,
$c= 3.7088\ \rm \AA$ for FePt \cite{alsaad2020,sawada2023} and
$a=b=3.8564\ \rm \AA $, $c= 3.7400\ \rm \AA$ for FePd. {
As shown by Laughlin \et \cite{laughlin2005}, this conventional unit
  cell can be further reduced, so the primitive cell is a
  base-centered tetragonal cell\cite{laughlin2005}, where the
  in-plane lattice constant is $a_t=a/\sqrt{2}$, Fe takes (0,0,0) and
  Pt/Pd takes $(\frac{1}{2},\frac{1}{2},\frac{1}{2})$. All our
  calculations use this primitive cell. We should point out an error
  in a prior publication by Ke \cite{ke2019}, where the primitive cell
  is mischaracterized as a body-centered tetragonal lattice.

We use a dense $\bk$ mesh of $25\times 25\times 18$, and set the
quantization axis along the $z$ axis.  The spin moment for FePt/FePd
is 3.26780/3.31417 $\mu_B$, which agree with those prior studies
\cite{sun2006,lu2010,sternik2015,marciniak2023} under the same
condition.  We carry out two separate calculations with and without
spin-orbit coupling (SOC) for the same $\bk$ mesh and same
functional. Then we compute the total energy difference between these
two cases, $\Delta E({\rm FePt})= E_{\rm soc}- E_{\rm nosoc}=-1.23$
eV, while $\Delta E({\rm FePd})= E_{\rm soc}- E_{\rm nosoc}=-0.136$
eV. This shows that the $5d$-Pt has a much stronger SOC than the
$4d$-Pd. This has an important consequence as seen below.  Figure
\ref{fig2}(a) is our band structure of FePt along seven high symmetry
lines. Consistent with the prior study \cite{ke2019}, FePt features
multiple bands crossing the Fermi level (at 0 eV), which opens many
channels for laser excitation.  Figure \ref{fig2}(b) is the density of
states (DOS) $\sigma\gc{0}$ in the ground state
\cite{sun2006,ueda2016}, where occupied states are between the Fermi
energy (set at 0 eV) and $-5$ eV, and are integrated to 18 electrons,
since Fe has $3d^64s^2$ and Pt has $5d^96s^1$ valence electrons. Note
that the DOS for the spin minority states is plotted on the negative
axis.  The band structure of FePd (Fig. \ref{fig2}(c)) is remarkably
similar to that of FePt between $-10$ and 5 eV (see
Fig. \ref{fig2}(a). The major difference is that FePd starts at $-8$
eV, while FePt at $-10$ eV. As a result, their DOS is also similar
(compare Figs. \ref{fig2}(b) and (d)). These similarities represent a
stringent test for our theory.

\section{Nonlinear optical quantum theory for spin response}

Charge dynamics in a material is different from spin dynamics
\cite{prb98} because their respective operators are different. Charge
dynamics is characterized by the electric polarization $\bP$, a polar
vector, defined as $\bP=\frac{1}{V}{\rm Tr}[\rho\bD]$, where Tr is a
trace, $\bD$ is the dipole moment, $V$ is the volume of the sample,
and $\rho$ is the density matrix.  Spin dynamics is characterized by
the magnetization $\bM$, an axial vector \cite{nc18,prb24b} defined
as $\bM=\frac{1}{V}{\rm Tr}[\rho\bS]$, where $\bS$ is the spin.  FePt
and FePd are centrosymmetric, and have inversion symmetry ${\cal I}$.
Under the inversion symmetry, $\bD\rightarrow -\bD$ but
$\bS\rightarrow \bS$. Figure \ref{fig1}(b) shows this difference.
$\rho$ does not have a simple expression under ${\cal I}$, but its
$n$th order\cite{mukamel} $\rho\gc{n}\propto \bD^n$, so $\bP\propto
\bD^{n+1}$.  It is helpful to examine a few lower orders. If $n=1$,
under ${\cal I}$, ${\cal I}\bP\gc{1} =\bP\gc{1}$, so $\bP\gc{1}\ne0$,
but ${\cal I}\bM\gc{1} =-\bM\gc{1}$, so $\bM\gc{1}=0$. If $n=2$,
${\cal I}\bP\gc{2} =-\bP\gc{2}$, so $\bP\gc{2}=0$, but ${\cal
  I}\bM\gc{2} =\bM\gc{2}$, so $\bM\gc{2}\ne 0$.  Therefore,
$\bM\gc{2}$ is the lowest possible magnetization for a centrosymmetric
system, consistent with our prior numerical results \cite{nc18} but
proved here in a much simpler way than done before \cite{mrudul2024}.
This underlines a fundamental difference between the traditional
nonlinear optical and magnetic responses, and will guide us through
all the following presentations.


\newcommand{\cP}{\bp}
\newcommand{\CP}{{\cal P}_l}
\newcommand{\CQ}{{\cal Q}_l}

\subsection{Second-order density matrix}

We employ the standard perturbation theory and expand the density
matrix $\rho=\rho\gc{0}+\rho\gc{1}+\rho\gc{2}+\cdots $, and then get a
hierarchy of equations for each order $n$.  The $n$th order
$\rho\gc{n}$ depends on the $(n-1)$th order density matrix. We start
with the first-order time-dependent Liouville equation \cite{jap25}
\be i\hbar
\dot{\rho}\gc{1}=[H_0,\rho\gc{1}]+[H\gd{a}{I},\rho\gc{0}],\label{eq1}
\ee where $\rho\gc{0}$ and $\rho\gc{1}$ are the zeroth- and
first-order density operators, respectively, and the dot over
$\rho\gc{1}$ is the time derivative. {Here $H_0$ is the unperturbed
Hamiltonian and includes the effect of the exchange functionals}.  The interaction Hamiltonian is $H\gd{a}{I}=\frac{e \bp
  \cdot \bA\gc{a}(t)}{2m_e}$, where $\bp$ is the electron momentum and
$m_e$ is the electron's mass.  We choose a cw field with the vector
potential being $\bA\gc{a}(t)= \bA\gd{a}{0}(e^{i\omega_a t}
+e^{-i\omega_a t}$)/2, where $\omega_a$ is the carrier frequency of
the laser field $a$ and $\bA\gd{a}{0}$ is the field amplitude. The
light fluence is
$F\gc{a}=\frac{1}{2}n\epsilon_0c(A\gd{a}{0}\omega_a)^2T=\pi nc
\epsilon_0(A\gd{a}{0})^2 \omega_a $, where $T$ is the laser period,
$n$ is the index of refraction, and $c$ is the speed of light. Since
our materials are crystallines, all the physical observables are
labeled by crystal momentum $\bk$, but for brevity, we hide it from
all the quantities below.

Introducing two band states $|n\ra$ and $|m\ra$ of $H_0$ allows us to
cast the density operator in Eq. \ref{eq1} into a
matrix form. Then we integrate over time to find $\rho\gc{1}$ as
\begin{widetext}
\be
\rho\gc{1}(n,m)=\frac{e\bA\gd{a}{0}\cdot \bp(n,m)}{2\hbar
  m_e}(\rho\gc{0}(n)-\rho \gc{0}(m)) \left (
\frac{e^{-i\omega_at}}{\omega_{nm}-\omega_a-i\Gamma_{nm}}
+\frac{e^{i\omega_at}}{\omega_{nm}+\omega_a-i\Gamma_{nm}} \right ),
\label{eq2}
\ee \end{widetext} where $\rho\gc{0}(n)$ is a shorthand notation of
$\rho\gc{0}(n,n)$, $\omega_{nm}=(E_n-E_m)/\hbar$, $E_n$ is the band
energy, $\Gamma_{nm}$ is a lifetime broadening, and $\bp(n,m)$ is the
momentum matrix element between bands $n$ and $m$.  Different from the
traditional nonlinear optics treatment \cite{boyd}, Eq. \ref{eq2}
includes both the resonant and off-resonant terms (the first and
second terms in Eq. \ref{eq2}), so $\rho\gc{1}(n,m)$ is Hermitian,
i.e., $\rho\gc{1}(n,m)=\rho\gc{1}^*(m,n)$, and can be used to compute
the spin moment change $m\gc{1}={\rm
  Tr}(\rho\gc{1}S_z)=\sum_{n,m}(\rho\gc{1}(n,m)S_z(m,n) $, where
$S_z(n,m)$ is the spin matrix.  {Equation \ref{eq2}
  further reveals that if $n=m$, $\rho\gc{1}(n,n)=0$, then only the
  off-diagonal first-order density matrix elements are nonzero. To
  have a nonzero $m\gc{1}$, $S_z(m,n)$ must have off-diagonal
  elements, i.e., the spin symmetry broken.  We note in passing that
  since our calculation always contains the spin-orbit coupling,
  density matrix elements are represented in the spin-mixed states,
  where the spin index cannot be used any longer.  }





With the first-order density matrix in hand, we are ready to work out
the second-order density matrix. We introduce a second light field
with the vector potential $\bA\gc{b}(t)= \bA\gc{b}(e^{i\omega_{b} t}
+e^{-i\omega_{b} t}$)/2, where $\omega_{b}$ is its carrier frequency.
The second-order $\rho\gc{2}$ obeys the second-order time-dependent
Liouville equation \be i\hbar
\dot{\rho}\gc{2}=[H_0,\rho\gc{2}]+[H\gd{b}{I},\rho\gc{1}]. \label{liou}
\ee We multiply both sides by band $\la n|$ from the left and band
$|m\ra$ from the right, and use the completeness relation \cite{ourqm}
to find
\begin{widetext}
\ba i\hbar
\dot{\rho}\gc{2}(n,m)&=&\la n|[H_0,\rho\gc{2}]|m\ra +\la
n|[H\gd{b}{I},\rho\gc{1}]|m\ra \nonumber
\\ &=&(E_n-E_m)\rho\gc{2}(n,m)+\sum_l\la n|H\gd{b}{I}|l\ra\la
l|\rho\gc{1}|m\ra -\la n|\rho\gc{1}|l\ra\la l|H\gd{b}{I}|m\ra\nonumber
\\ &=&(E_n-E_m)\rho\gc{2}(n,m)+\sum_lH\gd{b}{I}(n,l)\rho\gc{1}(l,m)
-\rho\gc{1}(n,l)H\gd{b}{I}(l,m).
\label{eq3}
\ea \end{widetext}
A lengthy but straightforward calculation (see Appendix A)
yields the second-order density matrix $\rho_{ab}^{(2)}(n,m)$ between
two band states $|n\ra$ and $|m\ra$, \small
\begin{widetext}
  \ba
\rho_{ab}^{(2)}(n,m)&=&\frac{e^2}{4m_e^2\hbar^2}\times \nonumber \\ 
&\sum_l&[\overbrace{\CQ(n,m,\omega_a,\omega_b)
  +\CQ^*(m,n,\omega_a,\omega_b)}^{\rm SFG_1}\nonumber
+
\overbrace{
\CQ(n,m,-\omega_a,-\omega_b) +
\CQ^*(m,n,-\omega_a,-\omega_b)}^{\rm SFG_2}  \\
  &+&\underbrace{\CQ(n,m,\omega_a,-\omega_b)+
  \CQ^*(m,n,\omega_a,-\omega_b)}_{\rm DFG_1} 
+\underbrace{\CQ(n,m,-\omega_a,\omega_b) +
  \CQ^*(m,n,-\omega_a,\omega_b)}_{\rm DFG_2}],
\label{main1}
\ea
\normalsize
where
{the first four terms are from the sum frequency
generation (SFG),  and the last four terms are from the
difference frequency generation (DFG),
$e$ is the
elementary charge, $\hbar$ is the reduced Planck constant, and the
summation over the crystal momentum $\bk$ is implied. These frequency
generation terms are identified through the frequency variables, i.e.,
$\omega_a$ and $\omega_b$
in $\CQ$. If $\omega_a$ and $\omega_b$ have the same sign, we have a
sum frequency generation; otherwise a difference frequency
generation. There are four possible combinations of $\pm\omega_a\pm\omega_b$.}  
$\CQ$ itself is 
given by, 
\be \CQ(n,m,\omega_a,\omega_b)=\frac{ [\bA\gc{a}\cdot
    \cP(n,l)][\bA\gc{b}\cdot \cP(l,m)] [\rho\gc{0}(n)-\rho\gc{0}(l)]
  e^{i(\omega_a+\omega_b)t}}
    {(\omega_{nm}+\omega_a+\omega_b-i\Gamma_{nm})(\omega_{nl}+\omega_a-i\Gamma_{nl})},
    \label{main2}  \ee
\end{widetext}
where $\cP(n,l)=\la n|\hp|l\ra$ is the momentum matrix element between
bands $|n\ra$ and $|l\ra$, $\omega_{nm}=(E_n-E_m)/\hbar$, $E_n$ is the
band energy, $\rho\gc{0}(l)$ is the ground-state occupation of band
$l$, and $t$ is the time. {All $\Gamma$'s are
  broadening, and they represent the disorder in the sample,
  electron-phonon sccattering and other scattering processes that are
  not explicitly treated in our theory.}  If we compare
Eq. \ref{main1} with Eq. 3.6.7 on page 135 of \cite{boyd}, we see that
his equation only contains $\CQ(n,m,-\omega_a,-\omega_b)$ and
$\CQ^*(m,n,\omega_a,\omega_b)$. One can easily verify that
$\rho_{ab}^{(2)}(n,m)$ with only $\CQ(n,m,-\omega_a,-\omega_b)$ and
$\CQ^*(m,n,\omega_a,\omega_b)$ is not even Hermitian, so
$\rho_{ab}^{(2)}(n,m)$ cannot be used to compute the spin moment. By
contrast, our $\rho_{ab}^{(2)}(n,m)$ is Hermitian, which ensures the
second-order spin moment $m\gd{2}{ab}={\rm Tr}[\rho\gd{2}{ab}S_z]$ to
be real.  So our theory represents a fundamental departure from the
traditional NLO theory.  {Since the third order is
  zero, i. e., $m\gc{3}=0$, the truncation at $m\gc{2}$ is a good
  start and is adequate when the vector potential is below $0.01\rm
  Vfs/\AA$ as seen in our numerical calculation below
  (Fig. \ref{fig5}(f)).  On the other hand, the fourth order $m\gc{4}$
  is analytically difficult to compute since it contains 128 terms.
  Microscopically, Eq. \ref{main2} represents a process that the
  magnet is kept irradiated by a continuous light wave, so the
  magnetization of the magnet can be affected. This is the essence of
  our theory.  }





\subsection{Second-order density of states}

\newcommand{\hpp}{\hat{p}}

The second-order density matrix has been extensively studied in
conjugated polymers \cite{mukamel}, but in solids, it remains largely
unexplored, partly because the focus has often been on the
second-order susceptibility.  {A quick inspection of the
  diagonal elements in Eq. \ref{main1} reveals that in contrast to its
  first-order counterpart ($\rho\gc{1}(n,n)=0$ in Eq. \ref{eq2}), it
  is nonzero, i.e., $\rho\gc{2}(n,n)\ne 0$, and obeys the sum rule,
  $\sum_n\rho\gc{2}(n,n)=0$.  To see what it entails, we disperse it
  into the energy domain and introduce the second-order density of
  states as} \be \sigma\gd{2}{ab}(E;\omega_p,\omega_q)=
\sum_{n}\frac{\rho\gd{2}{ab}(n,n)}{E_{n}-E+i\gamma},\label{eq12} \ee
which is not a simple second-order derivative of the ground-state DOS.
{Rather, it represents how electrons are excited out of
  the Fermi sea under two light fields.  For this reason, it
depends on both the excited-state property of a material,
  and the laser parameters such as laser
  polarization, photon energy and vector potential amplitude. This is
  fundamentally different from the ground-state DOS.}

We consider that two light fields $a$ and $b$ are both polarized along
the $x$ axis, i.e., collinear configuration, and have the same photon
energy $h\nu$.  We fix the incident fluence at $F=10$ \mje, typically
used in experiments \cite{jpcm10}.  Since the vector potential $A_0$
is inversely proportional to $\hbar\omega$, at $\hbar\omega=1.6$ eV we
have $A_0=0.222\ \rm Vfs/\AA$.  The thin solid line in
Fig. \ref{fig3}(a) is our $\sigma\gd{2}{xx}(E)$ in FePt. Different
from the regular DOS (see Figs. \ref{fig2}(b) and (d)), for $E<E_F$,
$\sigma\gd{2}{xx}(E)$ is negative, but becomes positive if
$E>E_F$. This stems from the population difference
$\rho\gc{0}(n)-\rho\gc{0}(l)$ in Eq. \ref{main2}, where the Pauli
exclusion principle is realized through the permutation relation in
Eq. \ref{liou}. The population difference differs from zero only when
one band is occupied or partially occupied and the other band is
unoccupied or partially unoccupied. This makes sense physically as
during excitation the valence bands only lose electrons and the
conduction bands only receive electrons.  We find the trough at
$-0.64$ eV and the peak at 0.72 eV, which are intrinsic to FePt,
nearly independent of $h\nu$.  This shows that those states close to
the Fermi surface make significant contributions to laser excitation
in terms of intraband transitions \cite{jpcm23}.  Because of the sum
rule, the positive and the negative $\sigma\gd{2}{xx}$ cancel to zero.
{Increasing $h\nu$ to 2.0 eV reduces the amplitude of
  $\sigma\gd{2}{xx}(E)$ because the vector potential is inversely
  proportional to $h\nu$ (see the thick dashed line in
  Fig. \ref{fig3}(a))}.  Figure \ref{fig3}(c) shows that in FePd the
trough is moved to $-1.00$ eV and the peak is moved to 0.75 eV, also
independent of $h\nu$. If we compare the amplitudes of
$\sigma\gd{2}{xx}(E)$ between FePt and FePd, we notice that FePt has a
larger amplitude, because FePt has stronger momentum transition matrix
elements.


Using the cross-polarized light fields, one along the $x$-axis and the
other along the $y$-axis, alters the picture completely.  The dashed
lines in Figs. \ref{fig3}(a) and \ref{fig3}(c) are $\sigma\gd{2}{xy}$
for FePt and FePd, respectively.  They are significantly smaller than
$\sigma\gd{2}{xx}$.  This feature is generic, as we find same results
in other materials.  This difference can be traced back to
Eq. \ref{main2}. Under cross-polarization excitation, microscopically
it is the off-diagonal matrix elements $\cP_x(n,l)\cP_y(l,n)$ that
contribute to $\sigma\gd{2}{xy}$. In contrast to
$\cP_x(n,l)\cP_x(l,n)$ which is always positive,
$\cP_x(n,l)\cP_y(l,n)$ can change signs.  This explains why
$\sigma\gd{2}{xy}$ can be positive or negative in different energy
regimes. However, one thing remains: the peak and trough positions are
also independent of $h\nu$.

\subsection{Light-induced spin moment change}


It is a known experimental fact that under a weak laser excitation,
the spin moment reduction is linearly proportional to the incident
laser fluence. We can first check whether our theory is able to
reproduce this.  {As mentioned above, our
  light-induced second-order spin moment is defined as \be
  m\gd{2}{ab}={\rm Tr}[\rho\gd{2}{ab}S_z], \label{main3} \ee where
  $\rho\gd{2}{ab}$ is from Eq. \ref{main1}, $S_z$ is the spin matrix
  in the spin-mixed states, not diagonal in general, and $a$ and $b$
  denote the light polarizations.}  We see immediately that since
$\rho\gd{2}{ab}$ in Eq. \ref{main2} is proportional to $\bA\gc{a}
\bA\gc{b}$, the induced spin moment is proportional to the fluence.
Now here comes to the most difficult question: How can one be sure
that $\rho\gc{2}$ delivers a negative $ m\gc{2}$ to have
demagnetization?  Under the cw approximation,
  $\rho\gd{2}{ab}$ in Eqs. \ref{main1} and \ref{main2} is time
  dependent, as is $m\gd{2}{ab}$. To be definitive, we choose a time
  instant at $t=0$.  Figure \ref{fig3}(b) has six curves for FePt, but
  only three are visible with a good reason. The filled cirlces,
  partially hidden behind the empty circles, denote $m\gd{2}{xx}$
  which is negative and real across the entire region of $h\nu$, also
  true for FePd (see Fig. \ref{fig3}(d)).  The reason for a negative
  $m\gc{2}$ lies in how $\rho\gc{2}$ changes across the Fermi
  energy. As shown in Figs. \ref{fig3}(a) and \ref{fig3}(b), during
  excitation, the spin majority states lose electrons and the spin
  minority states gain electrons, so one has a negative spin moment, a
  net angular momentum loss.  This demonstrates for the first time
that our nonlinear optical theory is capable of describing the
demagnetization.  {Using a denser $\bk$ mesh of
  $34\times 34\times 25$ yields nearly identical results (the empty
  circles), which overlap with the filled circles obtained at
  $25\times 25\times 18$. This shows that our results are well
  converged.}  {We also employ the LDA functional
  instead of GGA used above. The thin star and empty box lines are
  $m\gd{2}{xx}(\rm LDA)$ and $m\gd{2}{xy}(\rm LDA)$. They again
  overlap with the above data strongly. This shows that different
  functionals do not produce a significant change.  Across the same
  energy regime investigated, FePd has a weaker $m\gd{2}{xx}$ of
  around $-0.02$ \ub\ than FePt of around $-0.1$ \ub.}

Changing the light polarization affects $m\gc{2}$ significantly.  In
FePt, when $h\nu$ is above 0.7 eV, $m\gd{2}{xx}$ is larger than
$m\gd{2}{xy}$ (empty boxes).  Just as in the inverse Faraday effect,
different from $m\gd{2}{xx}$, $m\gd{2}{xy}$ is helicity-dependent,
$m\gd{2}{xy}=-m\gd{2}{yx}$.  In the figure, we choose a negative
$m\gd{2}{xy}$, so we can compare it with $m\gd{2}{xx}$ easily.
Quantitatively, at $h\nu=1.6$ eV, $m\gd{2}{xx}$ reaches $-0.09344$
\ub, 13.5 times larger than $m\gd{2}{xy}$ of $-0.0069$ $\mu_B$. But
below 0.7 eV, $m\gd{2}{xy}$ is stronger. This shows that in the THz
regime, the cross-polarization is equally effective to the spin
change.  In FePd, the crossing point where $m\gd{2}{xy}$ is larger
than $m\gd{2}{xx}$ is at 1.3 eV. This is consistent with the
differences seen in $\sigma\gc{2}$ in Figs. \ref{fig3}(a) and
\ref{fig3}(c).  {To this end, we use the broadening
  $\Gamma=0.05$ Ry. When we reduce it to 0.03 Ry (empty diamonds in
  Fig. \ref{fig3}(c)), $m\gd{2}{xx}$ negatively increases as
  expected.}   {Our theory can be applied to other
  magnets as well.  Figures \ref{fig3}(e) and (f) show $m\gd{2}{xx}$
  and $m\gd{2}{xy}$ for bcc Fe and fcc Ni, respectively. The trend is
  very interesting. Fe has a stronger response than Ni, in both
  $m\gd{2}{xx}$ and $m\gd{2}{xy}$, consistent with their native spin
  moments. What is different from FePt and FePd is that
the  cross-polarized spin moment $m\gd{2}{xy}$ is in general stronger
  than collinearly-polarized spin moment $m\gd{2}{xx}$. This reflects
  that $m\gc{2}$ is very sensitive to the intrinsic material
  properties.}

\subsection{Frequency-generation-resolved partial spin moment}
\label{newsec}

{To understand what and how optical processes
  underline the spin change, we compute the partial second-order spin
  moment, \be m\gd{2}{\rm DFG/SFG}={\rm Tr}(\rho\gd{2}{\rm
    DFG/SFG}S_z), \label{main4} \ee where $\rho\gc{2}$ is from
  Eq. \ref{main1} and is now split into eight time-dependent
  terms. They form four groups: two subgroups of sum frequency
  generation (SFG$_1$ and SFG$_2$) and another two subgroups of
  difference frequency generation (DFG$_1$ and DFG$_2$).  Each
  subgroup in Eq. \ref{main1} contains two terms.  Using one term
  leads to a complex $m\gd{2}{\rm DFG/SFG}$, and the sum of both terms
  results in a real $m\gd{2}{\rm DFG/SFG}$. This highlights the
  importance of Hermitian in the density matrix.}  {We take $\hbar\omega_a=\hbar\omega_b=1.6$ eV as an example.  Figure
  \ref{fig4}(a) shows all SFG terms have very small $m\gd{2}{xx}$ and
  beat with time $t$ at the frequency of $2\omega$ {because of
    their large denominator and the phase factor}. By contrast,
  DFG$_1$ and DFG$_2$ are much larger and do not oscillate with time
  because the phase factor is zero. Their partial $m\gc{2}$ are not
  the same. When we sum them up, we obtain the total $m\gc{2}$ as
  shown in Fig. \ref{fig4}(b). The solid line is the result of FePt,
  where the total $m\gc{2}$ oscillates around a negative value. FePd
  has a smaller negative $m\gc{2}$ (dotted line).
  {There is no other major change.
    Therefore, even at a particular time instant, our
  theoretical result can be compared with the experiments with a
  finite pulse duration.} 
  To be sure that our
  theory does not only apply to FePt and FePd, we also compute
  $m\gc{2}$ for bcc Fe (dashed line) and fcc Ni (long-dashed line). A
  trend is found. Whenever its ground-state spin moment is smaller,
  $m\gc{2}$ in general is smaller.}  Table \ref{table1} lists
$m\gd{2}{\rm DFG_1}$ and $m\gd{2}{\rm DFG_2}$ separately. One sees
that $m\gd{2}{\rm DFG_1}$ and $m\gd{2}{\rm DFG_2}$ each are very
large, but they differ by a sign.  For FePd, $m\gd{2}{\rm
  DFG_1}=-0.30346$\ub, and $m\gd{2}{\rm DFG_2}=0.28868$\ub. The net
second-order spin moment is the competition between these two large
numbers.  This is true for each material that we investigated. In the
table, we also include fcc Ni and bcc Fe.  Physically, the DFG groups,
with zero frequency $\omega_p-\omega_q=0$, correspond to the optical
rectification/shift current in nonlinear optics \cite{shen,boyd}.  The
light-induced rectification manifests itself in the second-order spin
moment.

\section{Laser-induced ultrafast demagnetization}

Can all the predictions above be realized in real-time dynamic
simulations? {There have been several prior studies
  on FePt using time-dependent density functional theory (TDDFT)
  \cite{elliot2022,chen2019}, but their demagnetizaion has been
  plagued by spurious rapid oscillations \cite{sharma2022,mrudul2024},
  which are absent from the experimental results \cite{iihama2016}}.
We employ the time-dependent Liouville equation \cite{jpcm23} which
does not have this problem. We choose a laser pulse of 60-fs and
1.6-eV. The vector potential amplitude is $A_0=0.015\rm\ Vfs/\AA$,
corresponding to the fluence of 1.34 \mje.  {In
  order to appropriately describe the collective excitation of
  conduction electrons in metals, we include the intraband transitions
  as described in our prior publication \cite{jpcm23}, where a bracket
  energy $\delta$ is used.}  Figure \ref{fig5}(a) shows that upon
laser excitation, both FePt and FePd demagnetize quickly, where
$\Delta M$ is the spin moment change $\Delta M=M(t)-M_0$, and $M_0$ is
the initial spin moment.  Quantitatively, Fig.  \ref{fig5}(a) shows
$\frac{\Delta M}{M_0}({\rm FePt})=-26.8$\% for FePt and $\frac{\Delta
  M}{M_0}({\rm FePd})=-19.4$ \% for FePd.  So for the same set of
laser parameters, FePt demagnetizes more than FePd, by 1.38 times
(26.8/19.4). A similar ratio was obtained experimentally.  Iihama \et
\cite{iihama2016} used two comparable fluences 1.6 \mj for FePd and
1.4 \mj for FePt, and they found that FePt demagnetizes smoothly by
7.5\% and FePd by 5\%, or 1.5 times larger in FePt. This agreement is
encouraging, given that there are many differences between the
experiment and theory.
To show that the demagnetization difference between FePt and FePd is
intrinsic to the materials themselves, not related to a particular set
of laser parameters.  When we increase the photon energy to 2.0 eV, we
find that FePt still demagnetizes more than FePd (see
Fig. \ref{fig5}(b)). This remains true regardless of whether we
increase the laser pulse duration to 120 fs (Fig. \ref{fig5}(c)) or
increase the vector potential amplitude to $0.03\ \rm Vfs/\AA$
(Fig. \ref{fig5}(d)). {Quantitatively, our
  percentage differs from the experimental one since in our current
  study, we do not purposely tune our bracket energy $\delta$
  \cite{jpcm23} to match the experimental one, as our goal here is to
  verify our above analytical theory. Figure \ref{fig5}(e) shows that
  the amount of demagnetization increases quickly with $\delta$. To
  match the experimental one, we only use $\delta=0.6$ eV. All the
  above results are obtained with $\delta=1$ eV.  Second,
  experimentally it is difficult to have the same efficiency as the
  theory, because many factors such as the sample surface reflection
  and surface roughness may lead to a smaller demagnetization. Figure
  \ref{fig5}(f) is the dependence of the demagnetization on the laser
  vector potential.}

{Finally, we wish to investigate the photon-energy
  dependence of ultrafast demagnetization, though we have provided the
  dependence for the above analytical results in
  Fig. \ref{fig3}(a). Figure \ref{fig6}(a) shows that for the fixed
  fluence and duration, the amount of demagnetization weakly depends
  on the photon energy. The reason why we have to fix the fluence and
  laser duration is because the vector potential $A_0$ depends on both
  the laser fluence and duration. Another interesting topic is the
  effects of the exchange-correlation functional on
  demagnetization. Figure \ref{fig6}(b) shows that the LDA functional
  leads to a stronger demagnetization, partly because the electronic
  states under the LDA functional have stronger transition matrix
  elements, as it builds upon the free electron gas model.  We also
  investigate whether the sample orientation matters to
  demagnetization. In this case, we choose two directions, one along
  the [111] direction and the other along the [101] direction. We
  apply linearly polarized laser pulses along those two
  directions. Figure \ref{fig6}(c) shows their orientation dependence
  is rather weak, because FePt, although having a L1$_0$ structure, is
  still quite symmetric spatially.} Our numerical studies prove that
our nonlinear optical quantum theory for spin change agrees with our
real-time demagnetization simulation.

\section{Conclusion}

We have developed the first nonlinear optical quantum theory of
demagnetization. We start from the symmetry analysis and find that for
centrosymmetric systems, the second-order spin moment is the lowest
order.  Our theory has two features.  First, different from nonlinear
optical theory, all the terms, regardless of whether they are resonant
or off-resonant, must be included to ensure the Hermitian of the
density matrix. Specifically, we show that the difference frequency
generations (DFG) dominate over the sum frequency generations
(SFG). This is the manifestation of optical rectification in spin
moment change.  The competition between two DFG terms determines the
net spin moment change.  Second, it allows one to compute and compare
light-induced spin moment changes among different magnetic materials
at the first-principles level. We find that FePt demagnetizes more
than FePd, even though their crystal and electronic structures are
very similar. This is confirmed in our real-time simulation and the
experiment \cite{iihama2016}.  We expect that our finding will
motivate further experimental and theoretical studies in
femtomagnetism.

\acknowledgments GPZ was partly supported by the U.S.\ Department of
Energy under Contract No.\ DE-FG02-06ER46304.  The work was carried
out on Indiana State University's Quantum Cluster and High Performance
computers.

$^*$guo-ping.zhang@outlook.com.
https://orcid.org/0000-0002-1792-2701


\appendix

\section{Derivation of Eq. \ref{main1}}
\label{app}

\begin{widetext}

Here we provide additional details of our derivation of
Eq. \ref{main1}.  To simplify our expression, we introduce
$\rho\gc{2}(n,m)=e^{-i\omega_{nm}t} Q(n,m)$, where
$\omega_{nm}=(E_n-E_m)/\hbar$, and substitute it into Eq. \ref{eq3} to
find \ba i\hbar (-i\omega_{nm} e^{-i\omega_{nm}t}Q(n,m)+
e^{-i\omega_{nm}t}\dot{Q}(n,m))\nonumber
\\ =(E_n-E_m)e^{-i\omega_{nm}t} Q(n,m)
+\sum_l[H\gd{b}{I}(n,l)\rho\gc{1}(l,m)
  -\rho\gc{1}(n,l)H\gd{b}{I}(l,m)] \\ i\hbar
e^{-i\omega_{nm}t}\dot{Q}(n,m)=\sum_l[H\gd{b}{I}(n,l)\rho\gc{1}(l,m)
  -\rho\gc{1}(n,l)H\gd{b}{I}(l,m)]\\ \dot{Q}(n,m)=\frac{1}{i\hbar}e^{i\omega_{nm}t}
\sum_l[H\gd{b}{I}(n,l)\rho\gc{1}(l,m)
  -\rho\gc{1}(n,l)H\gd{b}{I}(l,m)],
\label{eq4}
\ea
where 
$H\gd{b}{I}(n,l)=\frac{e} {2m_e} \bp(n,l) \cdot
\bA\gd{b}{0}(e^{i\omega_b t} +e^{-i\omega_b t})$
and
$H\gd{b}{I}(l,m)=\frac{e} {2m_e} \bp(l,m) \cdot
\bA\gd{b}{0}(e^{i\omega_b t} +e^{-i\omega_b t})$. Now we substitute
Eq. \ref{eq2} into the first term of 
Eq. \ref{eq4} to obtain
\ba
\frac{e^{i\omega_{nm}t}}{i\hbar}
H\gd{b}{I}(n,l)\rho\gc{1}(l,m)=\frac{e^{i\omega_{nm}t}}{i\hbar}
\frac{e} {2m_e} \bp(n,l) \cdot
\bA\gd{b}{0}(e^{i\omega_b t} +e^{-i\omega_b t})\nonumber\\
\times \frac{e\bA\gd{a}{0}\cdot \bp(l,m)}{2\hbar
  m_e}
(\rho\gc{0}(l)-\rho \gc{0}(m)) \left (
\frac{e^{i\omega_at}}{\omega_{lm}+\omega_a-i\Gamma_{lm}}
+\frac{e^{-i\omega_at}}{\omega_{lm}-\omega_a-i\Gamma_{lm}} \right )\\
=\frac{e^2 [\bp(n,l) \cdot
\bA\gd{b}{0}][
\bA\gd{a}{0}\cdot \bp(l,m)
][\rho\gc{0}(l)-\rho \gc{0}(m)]
}{4i\hbar^2 m_e^2}\nonumber\\
\times e^{i\omega_{nm}t} (e^{i\omega_b t} +e^{-i\omega_b t})
 \left (
\frac{e^{i\omega_at}}{\omega_{lm}+\omega_a-i\Gamma_{lm}}
+\frac{e^{-i\omega_at}}{\omega_{lm}-\omega_a-i\Gamma_{lm}} \right ). \label{eq5}
\ea
We focus on the second line of Eq. \ref{eq5} and  we multiply them out
to have
\ba
 (e^{i(\omega_b+\omega_{nm})t} +e^{-i(\omega_b-\omega_{nm}) t})
 \left (
\frac{e^{i\omega_at}}{\omega_{lm}+\omega_a-i\Gamma_{lm}}
+\frac{e^{-i\omega_at}}{\omega_{lm}-\omega_a-i\Gamma_{lm}} \right
) \nonumber \\
=
\frac{e^{i(\omega_a+\omega_b+\omega_{nm})t}}{\omega_{lm}+\omega_a-i\Gamma_{lm}}
+\frac{e^{-i(\omega_a+\omega_b-\omega_{nm})t}}{\omega_{lm}-\omega_a-i\Gamma_{lm}}
+\frac{e^{i(-\omega_a+\omega_b+\omega_{nm})t}}{\omega_{lm}-\omega_a-i\Gamma_{lm}}
+\frac{e^{i(\omega_a-\omega_b+\omega_{nm})t}}{\omega_{lm}+\omega_a-i\Gamma_{lm}}.  \label{eq6}
\ea
Next, we integrate each term from $-\infty$ to $t$. We take the first
term in Eq. \ref{eq6} as an example
\be
\int_{t'=-\infty}^{t'=t} \frac{e^{i(\omega_a+\omega_b+\omega_{nm})t'}}{\omega_{lm}+\omega_a-i\Gamma_{lm}}= \frac{e^{i(\omega_a+\omega_b+\omega_{nm})t}}
{i(\omega_{nm}+\omega_a+\omega_b
-i\Gamma_{nm})
(\omega_{lm}+\omega_a-i\Gamma_{lm})}, \label{eq11}
\ee
where we have introduced the decaying factor $e^{\Gamma_{nm}t}$
 ($\Gamma_{nm}>0$) 
so the integral at
$t'=-\infty$ is zero. The remaining terms are obtained by changing
$(\omega_a,\omega_b)$ to $(-\omega_a,-\omega_b)$,
$(-\omega_a,\omega_b)$, and    $(\omega_a,-\omega_b)$, respectively,
so the first
term in  $Q(n,m)$ in Eq. \ref{eq4}
is 
\ba
\sum_l
\frac{e^2 [\bp(n,l) \cdot
\bA\gd{b}{0}][\bA\gd{a}{0}\cdot \bp(l,m)
][\rho\gc{0}(l)-\rho \gc{0}(m)]}{4i\hbar^2 m_e^2}\nonumber \\
\times \left (
\frac{e^{i(\omega_a+\omega_b+\omega_{nm})t}}
{i(\omega_{nm}+\omega_a+\omega_b
-i\Gamma_{nm})
(\omega_{lm}+\omega_a-i\Gamma_{lm})}+
 \frac{e^{-i(\omega_a+\omega_b-\omega_{nm})t}}
{i(\omega_{nm}-\omega_a-\omega_b
-i\Gamma_{nm})
(\omega_{lm}-\omega_a-i\Gamma_{lm})}\nonumber \right  .\\
\left .
+ \frac{e^{i(-\omega_a+\omega_b+\omega_{nm})t}}
{i(\omega_{nm}-\omega_a+\omega_b
-i\Gamma_{nm})
(\omega_{lm}-\omega_a-i\Gamma_{lm})}+
 \frac{e^{i(\omega_a-\omega_b+\omega_{nm})t}}
{i(\omega_{nm}+\omega_a-\omega_b
-i\Gamma_{nm})
(\omega_{lm}+\omega_a-i\Gamma_{lm})}\nonumber\right )\\
=-\frac{e^2}{4\hbar^2m_e^2}\sum_l[\bp(n,l) \cdot
\bA\gd{b}{0}][\bA\gd{a}{0}\cdot \bp(l,m)
][\rho\gc{0}(l)-\rho \gc{0}(m)]\nonumber \\
\times \left (
\frac{e^{i(\omega_a+\omega_b+\omega_{nm})t}}
{(\omega_{nm}+\omega_a+\omega_b
-i\Gamma_{nm})
(\omega_{lm}+\omega_a-i\Gamma_{lm})}+
 \frac{e^{-i(\omega_a+\omega_b-\omega_{nm})t}}
{(\omega_{nm}-\omega_a-\omega_b
-i\Gamma_{nm})
(\omega_{lm}-\omega_a-i\Gamma_{lm})}\nonumber \right  . \nonumber \\
\left .
+ \frac{e^{i(-\omega_a+\omega_b+\omega_{nm})t}}
{(\omega_{nm}-\omega_a+\omega_b
-i\Gamma_{nm})
(\omega_{lm}-\omega_a-i\Gamma_{lm})}+
 \frac{e^{i(\omega_a-\omega_b+\omega_{nm})t}}
{(\omega_{nm}+\omega_a-\omega_b
-i\Gamma_{nm})
(\omega_{lm}+\omega_a-i\Gamma_{lm})}\nonumber\right )\nonumber
\ea

The second term in $Q$ in Eq. \ref{eq4}  can
 be worked out similarly,
\ba
-\frac{e^{i\omega_{nm}t}}{i\hbar}
H_I(l,m)\rho\gc{1}(n,l)=-\frac{e^{i\omega_{nm}t}}{i\hbar}
\frac{e} {2m_e} \bp(l,m) \cdot
\bA\gd{b}{0}(e^{i\omega_b t} +e^{-i\omega_b t})\nonumber\\
\times \frac{e\bA\gd{a}{0}\cdot \bp(n,l)}{2\hbar
  m_e}
(\rho\gc{0}(n)-\rho \gc{0}(l)) \left (
\frac{e^{i\omega_at}}{\omega_{nl}+\omega_a-i\Gamma_{nl}}
+\frac{e^{-i\omega_at}}{\omega_{nl}-\omega_a-i\Gamma_{nl}} \right )\\
=-\frac{e^2 [\bp(l,m) \cdot
\bA\gd{b}{0}][
\bA\gd{a}{0}\cdot \bp(n,l)
][\rho\gc{0}(n)-\rho \gc{0}(l)]
}{4i\hbar^2 m_e^2}\nonumber\\
\times e^{i\omega_{nm}t} (e^{i\omega_b t} +e^{-i\omega_b t})
 \left (
\frac{e^{i\omega_at}}{\omega_{nl}+\omega_a-i\Gamma_{nl}}
+\frac{e^{-i\omega_at}}{\omega_{nl}-\omega_a-i\Gamma_{nl}} \right ).
\label{eq20}
\ea
The second line of Eq. \ref{eq20} also contains four terms as
\ba
 (e^{i(\omega_b+\omega_{nm})t} +e^{-i(\omega_b-\omega_{nm}) t})
 \left (
\frac{e^{i\omega_at}}{\omega_{nl}+\omega_a-i\Gamma_{nl}}
+\frac{e^{-i\omega_at}}{\omega_{nl}-\omega_a-i\Gamma_{nl}} \right
) \nonumber \\
=
\frac{e^{i(\omega_a+\omega_b+\omega_{nm})t}}{\omega_{nl}+\omega_a-i\Gamma_{nl}}
+\frac{e^{-i(\omega_a+\omega_b-\omega_{nm})t}}{\omega_{nl}-\omega_a-i\Gamma_{nl}}
+\frac{e^{i(-\omega_a+\omega_b+\omega_{nm})t}}{\omega_{nl}-\omega_a-i\Gamma_{nl}}
+\frac{e^{i(\omega_a-\omega_b+\omega_{nm})t}}{\omega_{nl}+\omega_a-i\Gamma_{nl}},   \label{eq21}
\ea
whose respective time-integrals are
\ba
\left (
\frac{e^{i(\omega_a+\omega_b+\omega_{nm})t}}
{i(\omega_{nm}+\omega_a+\omega_b
-i\Gamma_{nm})
(\omega_{nl}+\omega_a-i\Gamma_{nl})}+
 \frac{e^{-i(\omega_a+\omega_b-\omega_{nm})t}}
{i(\omega_{nm}-\omega_a-\omega_b
-i\Gamma_{nm})
(\omega_{nl}-\omega_a-i\Gamma_{nl})}\nonumber \right  .\\
\left .
+ \frac{e^{i(-\omega_a+\omega_b+\omega_{nm})t}}
{i(\omega_{nm}-\omega_a+\omega_b
-i\Gamma_{nm})
(\omega_{nl}-\omega_a-i\Gamma_{nl})}+
 \frac{e^{i(\omega_a-\omega_b+\omega_{nm})t}}
{i(\omega_{nm}+\omega_a-\omega_b
-i\Gamma_{nm})
(\omega_{nl}+\omega_a-i\Gamma_{nl})}\nonumber\right ).
\ea
We then multiply it by the coefficient
$-\frac{e^2 [\bp(l,m) \cdot
\bA\gd{b}{0}][
\bA\gd{a}{0}\cdot \bp(n,l)
][\rho\gc{0}(n)-\rho \gc{0}(l)]
}{4i\hbar^2 m_e^2}$ to find
\ba
\frac{e^2 [\bp(l,m) \cdot
\bA\gd{b}{0}][
\bA\gd{a}{0}\cdot \bp(n,l)
][\rho\gc{0}(n)-\rho \gc{0}(l)]
}{4\hbar^2 m_e^2}\nonumber \\
\times \left (
\frac{e^{i(\omega_a+\omega_b+\omega_{nm})t}}
{(\omega_{nm}+\omega_a+\omega_b
-i\Gamma_{nm})
(\omega_{nl}+\omega_a-i\Gamma_{nl})}+
 \frac{e^{-i(\omega_a+\omega_b-\omega_{nm})t}}
{(\omega_{nm}-\omega_a-\omega_b
-i\Gamma_{nm})
(\omega_{nl}-\omega_a-i\Gamma_{nl})}\nonumber \right  .\\
\left .
+ \frac{e^{i(-\omega_a+\omega_b+\omega_{nm})t}}
{(\omega_{nm}-\omega_a+\omega_b
-i\Gamma_{nm})
(\omega_{nl}-\omega_a-i\Gamma_{nl})}+
 \frac{e^{i(\omega_a-\omega_b+\omega_{nm})t}}
{(\omega_{nm}+\omega_a-\omega_b
-i\Gamma_{nm})
(\omega_{nl}+\omega_a-i\Gamma_{nl})}\nonumber\right ).
\ea

Since $\rho\gc{2}(n,m)=e^{-i\omega_{nm}t}Q(n,m)$, all we need to do is
to remove $e^{i\omega_{nm}t}$ from the above expressions to get 
\ba
\rho\gc{2}(n,m)=\frac{e^2}{4\hbar^2m_e^2}\sum_l[\bp(n,l) \cdot
\bA\gd{a}{0}][\bA\gd{b}{0}\cdot \bp(l,m)
][\rho\gc{0}(n)-\rho \gc{0}(l)]\nonumber \\
\times \left ( \underbrace{
\frac{e^{i(\omega_a+\omega_b)t}}{(\omega_{nm}+\omega_a+\omega_b-i\Gamma_{nm})(\omega_{nl}+\omega_a-i\Gamma_{nl})}}_{\CQ(n,m,\omega_a,\omega_b)} 
+\underbrace{\frac{e^{-i(\omega_a+\omega_b)t}}{(\omega_{nm}-\omega_a-\omega_b-i\Gamma_{nm})(\omega_{nl}-\omega_a-i\Gamma_{nl})}}_{\CQ(n,m,\omega_a,\omega_b)}\nonumber \right .\\
\left . \underbrace{\frac{e^{i(-\omega_a+\omega_b)t}}{(\omega_{nm}-\omega_a+\omega_b-i\Gamma_{nm})(\omega_{nl}-\omega_a-i\Gamma_{nl})}}_{\CQ(n,m,-\omega_a,\omega_b)} 
+\underbrace{\frac{e^{i(\omega_a-\omega_b)t}}{(\omega_{nm}+\omega_a-\omega_b-i\Gamma_{nm})(\omega_{nl}+\omega_a-i\Gamma_{nl})}}_{\CQ(n,m,\omega_a,-\omega_b)}\nonumber \right
) \\
-\frac{e^2}{4\hbar^2m_e^2}\sum_l[\bp(n,l) \cdot
\bA\gd{b}{0}][\bA\gd{a}{0}\cdot \bp(l,m)
][\rho\gc{0}(l)-\rho \gc{0}(m)]\nonumber \\
\times \left (
\underbrace{\frac{e^{i(\omega_a+\omega_b)t}}
{(\omega_{nm}+\omega_a+\omega_b
-i\Gamma_{nm})
(\omega_{lm}+\omega_a-i\Gamma_{lm})}}_{\CQ^*(m,n,-\omega_a,-\omega_b)}+
\underbrace{\frac{e^{-i(\omega_a+\omega_b)t}}
{(\omega_{nm}-\omega_a-\omega_b
-i\Gamma_{nm})
(\omega_{lm}-\omega_a-i\Gamma_{lm})}}_{\CQ^*(m,n,\omega_a,\omega_b)}\nonumber \right  . \nonumber \\
\left .
+ \underbrace{\frac{e^{i(-\omega_a+\omega_b)t}}
{(\omega_{nm}-\omega_a+\omega_b
-i\Gamma_{nm})
(\omega_{lm}-\omega_a-i\Gamma_{lm})}}_{\CQ^*(m,n,\omega_a,-\omega_b)}+
 \underbrace{\frac{e^{i(\omega_a-\omega_b)t}}
{(\omega_{nm}+\omega_a-\omega_b
-i\Gamma_{nm})
(\omega_{lm}+\omega_a-i\Gamma_{lm})}}_{\CQ^*(m,n,-\omega_a,\omega_b)}\nonumber\right ),\nonumber
\ea
where we identify each term with $\CQ$. Caution must be taken that the
actual $\CQ$ includes the coefficients (see Eq. \ref{main2}).

\end{widetext}

\begin{widetext}
\begin{table}
  \caption{
Table of symbols used in this paper. 
}
\begin{tabular}{cl}
  \hline\hline
Symbol & Meaning \\
\hline
$\bP, \bP^{(n)}$
 & electric polarization, $n$th order\\
$\bD
$ & dipole moment\\
$\rho$, $\rho^{(n)}$
 & density matrix, $n$th order \\
$\bM, \bM^{(n)}
$ & magnetization, $n$th order\\
$M(t),\Delta M$& time-dependent spin moment, its change\\
${\cal I}$ & inversion symmetry operator\\
$H_0, H_I$ & system Hamiltonian, interaction Hamiltonian
\\
$\bp(n,m)$ & momentum matrix element between bands $n$ and $m$\\
$F^{(a)}$ & fluence of light field $a$\\
$\bA^{(a)},\omega_a,\nu_a$ & vector potential, angular frequency and frequency of  field $a$\\
$\omega_{nm}$ & angular frequency difference between bands  $n$ and $m$\\
$\Gamma_{nm}$ & lifetime broadening difference for bands  $n$ and
$m$\\
$\CQ$& shorthand notation for sum and difference frequency generations in
Eq. \ref{main2}
\\
$\sigma^{(2)}_{ab}$ & second-order density of states for fields $a$ and
$b$ (Eq. \ref{eq12})\\
$m_{ab}^{(2)}$ & light-induced second-order spin moment fields $a$ and
$b$
\\
\hline
\end{tabular}
\label{table0}
  \end{table}
\end{widetext}

\begin{table}
  \caption{Partial second-order spin moment $m\gd{2}{\rm DFG}$
    due to the difference frequency generation in FePd, FePt, bcc Fe
    and fcc Ni. They are computed from $m\gd{2}{\rm DFG_{1}} = {\rm
      Tr}[\rho\gd{2}{\rm DFG_1} S_z]$ and $m\gd{2}{\rm DFG_{2}} ={\rm
      Tr}[\rho\gd{2}{\rm DFG_2} S_z] $, where $\rho\gd{2}{\rm DFG_1}$
    and $\rho\gd{2}{\rm DFG_2}$ are the third and fourth terms in
    Eq. \ref{main1}.  $m\gd{2}{xx}$ is the net second-order spin
    moment, $m\gd{2}{xx}=m\gd{2}{\rm DFG_{1}}+m\gd{2}{\rm DFG_{2}}$.
    $m\gc{0}$ is the ground-state spin moment. $m\gd{2}{xx}/m\gc{0}$
    is the percentage change. All SFG terms are small, so are not
    included.  Here the photon energy is $h\nu=1.6$ eV. All the
    results are calculated at $t=0$.}
\begin{tabular}{cccccc}
  \hline\hline
Material &  $m\gd{2}{\rm DFG_2}$(\ub) &$m\gd{2}{\rm DFG_1}$(\ub) &$m\gd{2}{xx}$(\ub)
&$m\gc{0}$(\ub) & $m\gd{2}{xx}/m\gc{0}(\%)$\\
\hline
FePd & $-0.30346$ & 0.28868 & $-0.01678$& 3.3142 & $-0.445$\\
FePt  & $-0.3420 $ & 0.24968 & $-0.09232$& 3.2678 & $-2.825$\\
bcc  Fe & $-0.14066 $ & 0.13682 & $-0.00384$& 2.1770 & $-0.176$\\
fcc Ni  & $-0.05052 $ & 0.04842 & $-0.00210$& 0.6389 & $-0.328$\\
\hline
\end{tabular}
\label{table1}
  \end{table}

\begin{figure}
\includegraphics[angle=0,width=0.8\columnwidth]{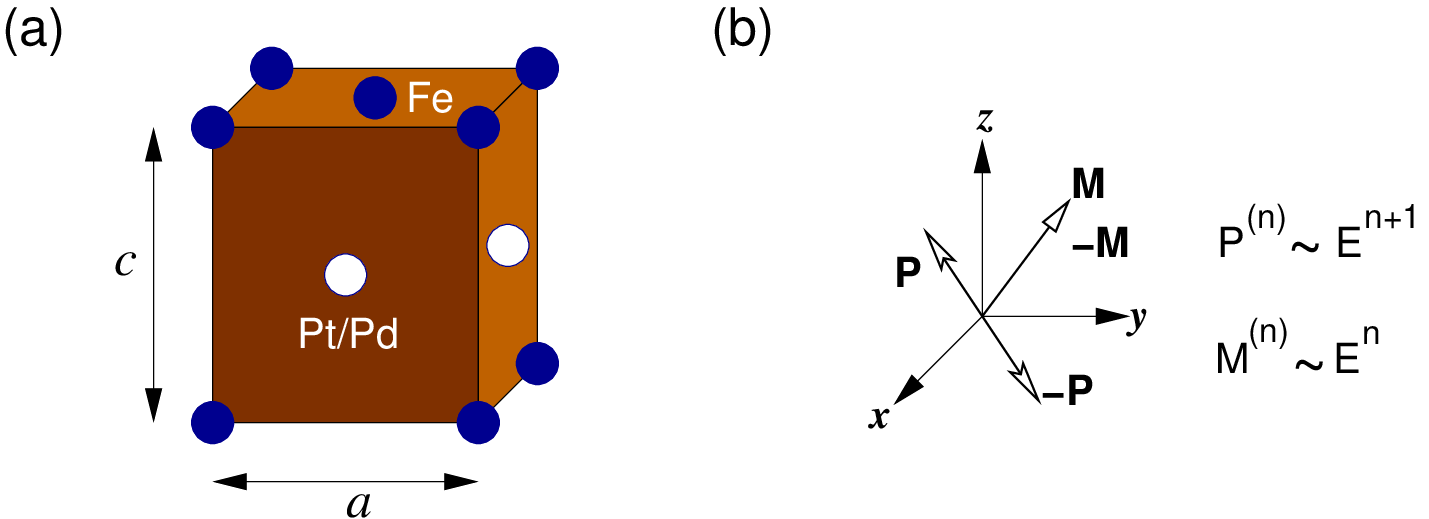}
\caption{ (a) FePt/FePd crystal structures. The filled circles are Fe
  atoms, while the unfilled ones are Pt/Pd.  (b) Symmetry difference
  between the polarization $\bP$ and magnetization $\bM$ determines
  how the $n$th-order polarization and magnetization depend on the
  external electric field $\bE$ differently.}
\label{fig1}
\end{figure}

\begin{figure}
\includegraphics[angle=0,width=0.8\columnwidth]{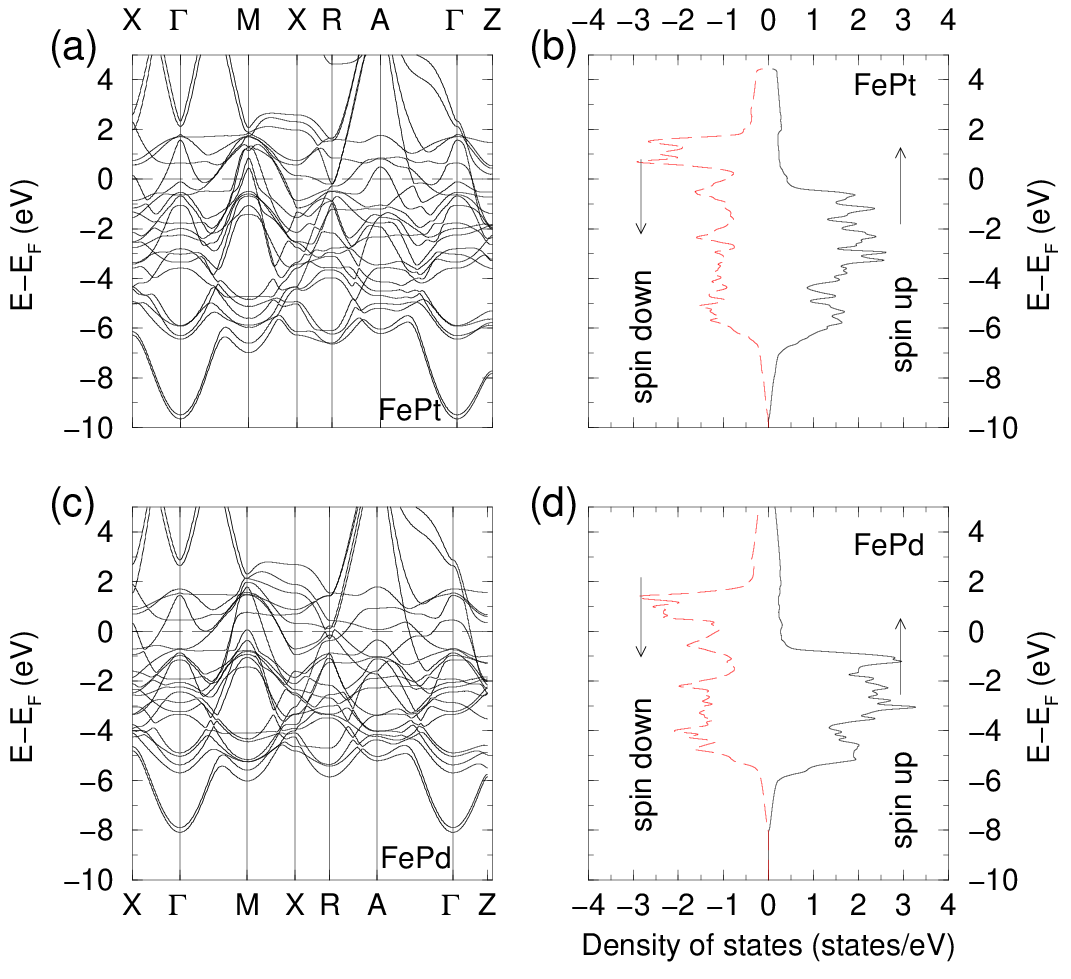}
\caption{ (a) Band structure of FePt. (b) Density of states of the
  ground state in FePt, where the solid line denotes the spin majority
  states and the dashed line the spin minority states plotted on the
  negative axis.  (c) Band structure of FePd. (d) Density of
  states of the ground state in FePd.  }
\label{fig2}
\end{figure}

\begin{figure}
\includegraphics[angle=0,width=0.8\columnwidth]{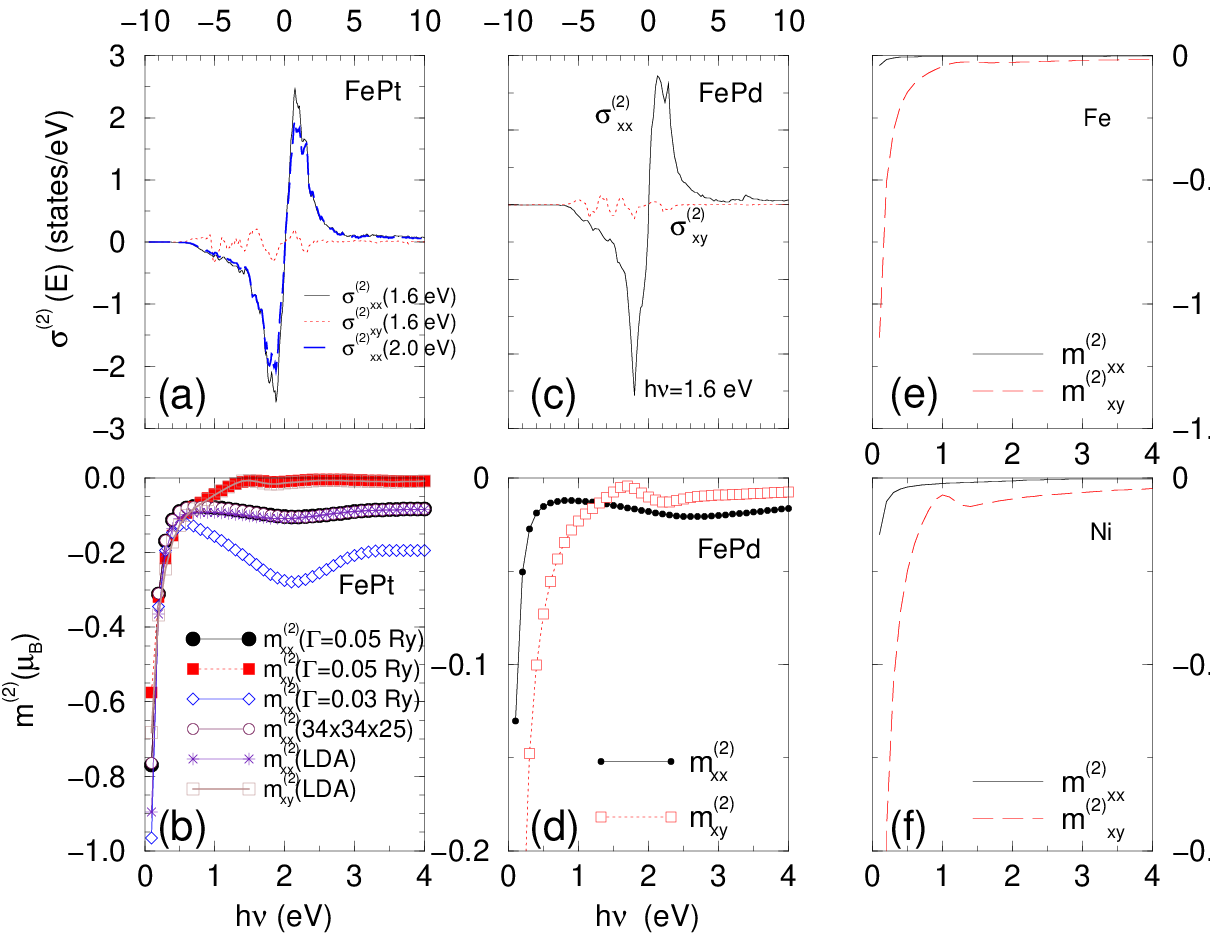}
\caption{(a) Second-order density of states in FePt.  The solid
  line denotes $\sigma\gd{2}{xx}$ with both laser polarizations along
  the $x$ axis, while the dotted line denotes $\sigma\gd{2}{xy}$ with
  cross-polarizations along the $x$ and $y$ axes.  Here the photon
  energy is $h\nu_a=h\nu_b=1.6$ eV.  The long-dashed line is
  $\sigma\gd{2}{xx}$ with $h\nu_a=h\nu_b=2.0$ eV.  (b) Second-order
  magnetic moment $m\gc{2}$ in FePt as a function of photon energy
  $h\nu$. The key feature is that except $m\gd{2}{xx}$ with a
  different broadening of $\Gamma=0.03$ Ry (empty diamonds), the
  convergences with the $\bk$ points and functionals are reached
  within 1-3\% as estimated from their overlaps. Here the filled
  circles denote $m\gd{2}{xx}$ and the filled boxes denote
  $m\gd{2}{xy}$ components, both with $\Gamma=0.05$ Ry.  The empty
  circles are $m\gd{2}{xx}$ with a larger $\bk$ mesh of $34\times
  34\times 25$. The thin star line and empty boxes are $m\gd{2}{xx}$
  and $m\gd{2}{xy}$ computed with LDA instead of GGA. All the
  calculations are done with one parameter changed, while the rest are
  fixed.  (c) Same as (a) but for FePd.  (d) Same as (c) but for
  FePd. (e) and (f) are $m\gd{2}{xx}$ and $m\gd{2}{xy}$ for bcc Fe and
  fcc Ni, respectively.  }
\label{fig3}
\end{figure}

 \begin{figure}

\includegraphics[angle=0,width=0.8\columnwidth]{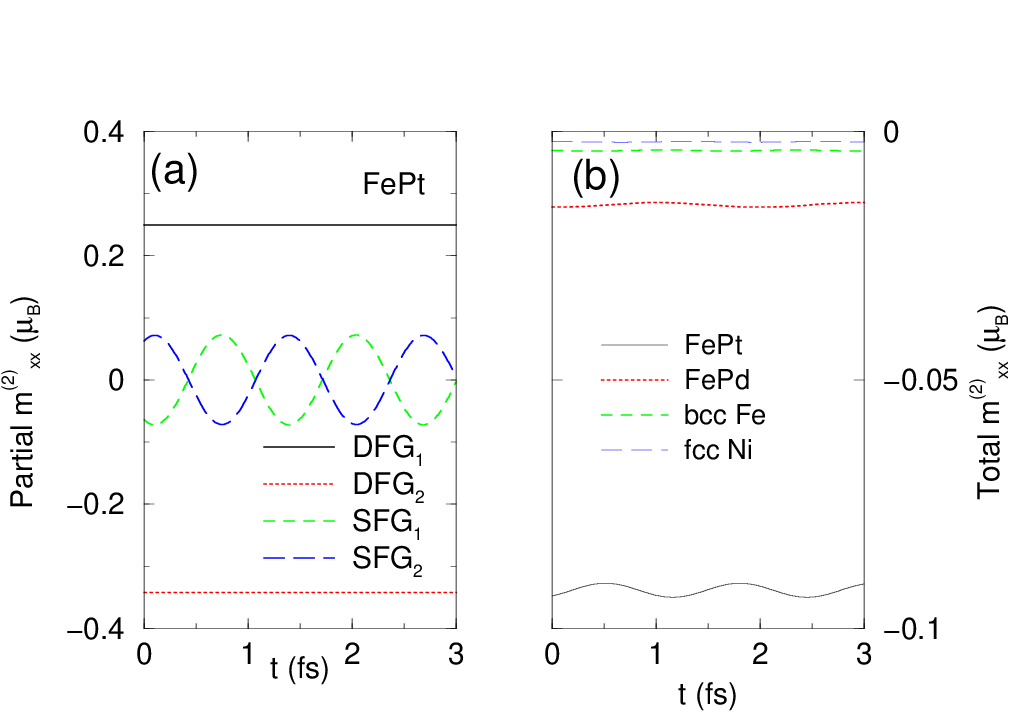}
\caption{(a) Contribution of the sum frequency [SFG$_1$ (dashed
    line), SFG$_2$ (long-dashed line)] and difference frequency
  generations [DFG$_1$ (solid line), DFG$_2$ (dotted line)] to
  $m\gc{2}$ as a function of time $t$. Here the photon energy  
is $h\nu_a=\hbar\nu_b=1.6$ eV.  (b) Total
  $m\gd{2}{xx}$  as a function of time $t$ for FePt (solid line), FePd
  (dotted line), bcc Fe (dashed line) and fcc Ni (long-dashed line).
 }
\label{fig4}
\end{figure}

\begin{figure}

\includegraphics[angle=0,width=0.8\columnwidth]{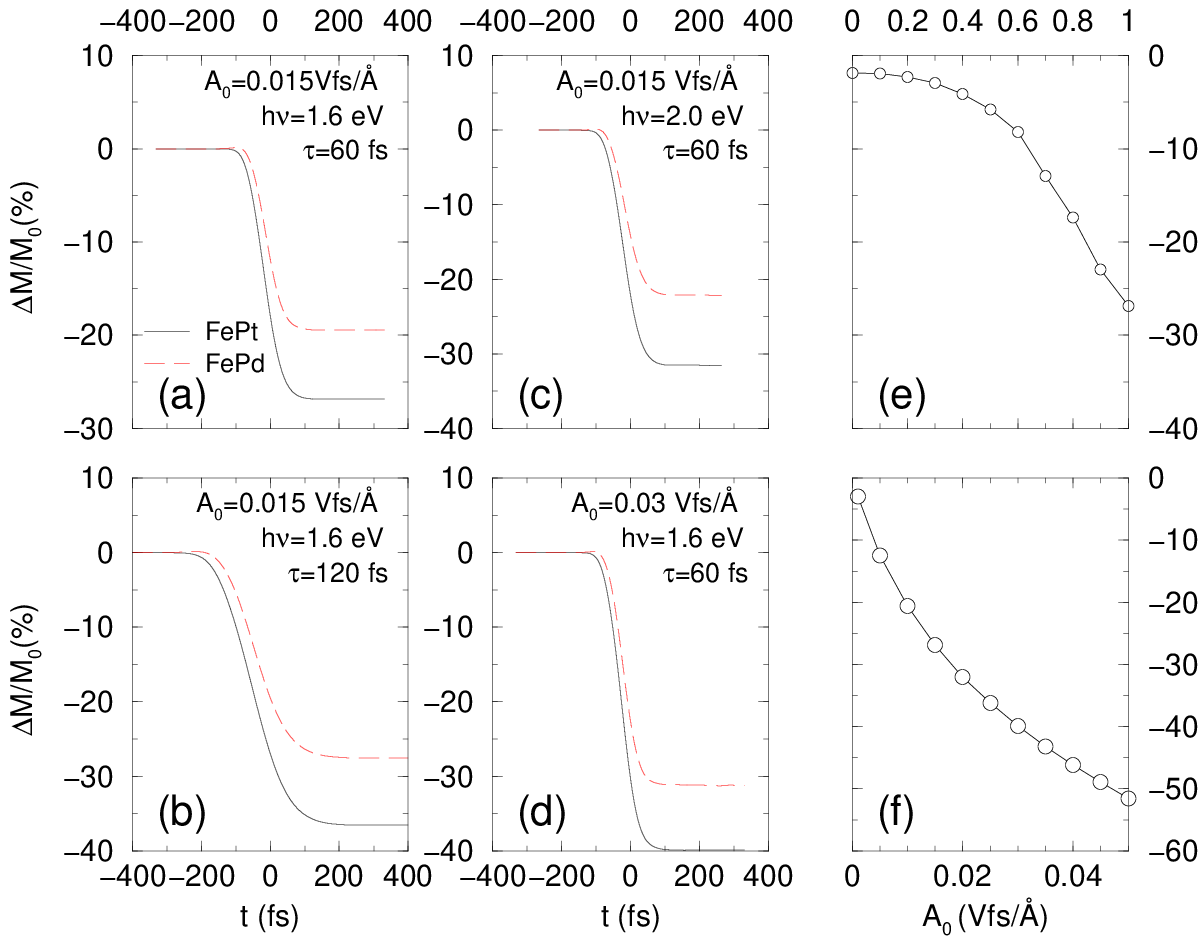}
\caption{Laser-parameter and intraband-bracket energy dependence
  of ultrafast demagnetization in FePt (black solid line) and FePd
  (red dashed line). (a) The laser photon energy is $h\nu=1.6$ eV,
  vector field potential is $A_0=0.015\ \rm Vfs/\AA, $ and pulse
  duration $\tau=60$ fs.  (b) Same as (a) but with the pulse duration
  $\tau=120$ fs. (c) Same as (a) but with $h\nu=2.0 $ eV.  (d) $A_0$
  is increased to 0.03 $\rm Vfs/\AA$. (e) The spin moment reduction in
  FePt as a function of the bracket energy $\delta$ which controls the
  contribution of the intraband transitions. (f) The spin moment
  reduction in FePt as a function of vector potential amplitude.  }
\label{fig5}
\end{figure}

 \begin{figure}

\includegraphics[angle=0,width=1\columnwidth]{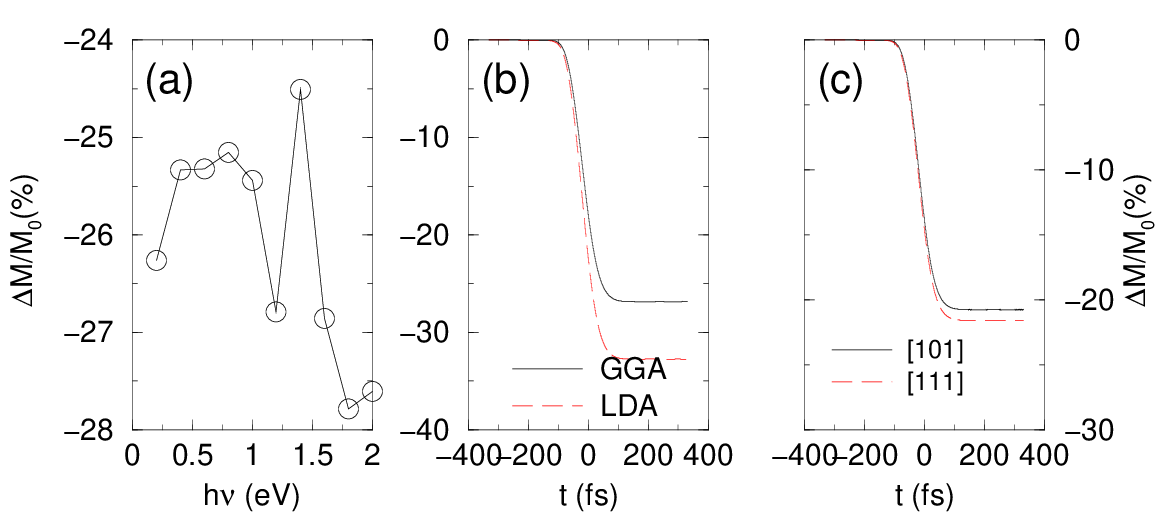}
\caption{(a) Photon energy dependence of demagnetization in
  FePt. Since our vector potential $A_0$ is photon-energy dependent,
  we fix the fluence at 1.34 \mj and pulse duration at $\tau=60$ fs.
  (b) Comparison of the spin moment reduction between the GGA (solid
  line) and LDA results (long-dashed line), where $h\nu=1.6$ eV,
  $\tau=60$ fs and $A_0=0.015\rm\ Vfs/\AA$.  Using the LDA functional
  produces a stronger demagnetization, but they agree within 10\%.
  (c) Orientation dependence. The solid line denotes the result with
  the laser linear polarization along the [101] direction, and the
  long-dashed line is for the [111] direction. We find their
  difference is small.  }
\label{fig6}
\end{figure}

\end{document}